**Why Your IT Project May Be Riskier than You Think**

**New research shows surprisingly high numbers of out-of-control tech projects — ones that can sink entire companies and careers.**

By

Bent Flyvbjerg and Alexander Budzier

BT Centre for Major Programme Management

Saïd Business School

University of Oxford

**Abstract**

Out-of-control information technology (IT) projects have ended the careers of top managers, such as EADS CEO Noël Forgeard and Levi Strauss' CIO David Bergen. Moreover, IT projects have brought down whole companies, like Kmart in the US and Auto Windscreen in the UK. Software and other IT is now such an integral part of most business processes and products that CEOs must know their IT risks, which are typically substantial – and overlooked. The analysis of a sample of 1,471 IT projects showed that the average cost overrun was 27% — but that figure masks a far more alarming "fat tail" risk. Fully one in six of the projects in the sample was a Black Swan, with a cost overrun of 200%, on average, and a schedule overrun of almost 70%. This highlights the true pitfall of IT change initiatives: It's not that they're particularly prone to high cost overruns *on average* – it is that there are a disproportionate number of Black Swans. By focusing on averages instead of the more damaging outliers, most managers and consultants have been missing the real risk in doing IT. In conclusion, the article outlines ideas as to what can be done to avoid Black Swans.

**Why Your IT Project May Be Riskier Than You Think**

To top managers at Levi Strauss, revamping the information technology system seemed like a good idea. The company had come a long way since its founding in the 19th century by a German-born dry-goods salesman: In 2003 it was a global corporation, with operations in more than 110 countries. But its IT network was antiquated, a balkanized mix of incompatible country-specific computer systems. So executives decided to migrate to a single SAP system and hired a team of Deloitte consultants to lead the effort. The risks seemed small: The proposed budget was less than $5 million. But very quickly all hell broke loose. One major customer, Walmart, required that the system interface with its supply chain management system, creating additional hurdles. Insufficient procedures for financial reporting and internal controls nearly forced Levi Strauss to restate quarterly and annual results. During the switchover, it was unable to fill orders and had to close its three U.S. distribution centers for a week. In the second quarter of 2008, the company took a $192.5 million charge against earnings to compensate for the botched project — and its chief information officer, David Bergen, was forced to resign.

A $5 million project that leads to an almost $200 million loss is a classic "Black Swan." The term was coined by our colleague Nassim Nicholas Taleb to describe high-impact events that are rare and unpredictable but in retrospect seem not so improbable. Indeed, what happened at Levi Strauss occurs all too often, and on a much larger scale. IT projects are now so big, and they touch so many aspects of an organization, that they pose a singular new risk. Mismanaged IT projects routinely cost the jobs of top managers, as happened to EADS CEO Noël Forgeard. They have sunk whole corporations. Even cities and nations are in peril. Months of relentless IT problems at Hong Kong's airport, including glitches in the flight information display system and the database for tracking cargo shipments, reportedly cost the economy $600 million in lost business in 1998 and 1999. The CEOs of companies undertaking significant IT projects should be acutely aware of the risks. It will be no surprise if a large, established company fails in the coming years because of an out-of-control IT project. In fact, the data suggest that one or more will.

We reached this bleak conclusion after conducting the largest global study ever of IT change initiatives. We examined 1,471 projects, comparing their budgets and estimated performance benefits with the actual costs and results. They ran the gamut from enterprise resource planning to management information and customer relationship management systems. Most, like the Levi Strauss project, incurred high expenses — the average cost was $167 million, the largest $33 billion — and many were expected to take several years. Our sample drew heavily on public agencies (92%) and U.S.-based projects (83%), but we found little difference between them and projects at the government agencies, private companies, and European organizations that made up the rest of our sample.

**The True IT Pitfall**

When we broke down the projects' cost overruns, what we found surprised us. The average overrun was 27% — but that figure masks a far more alarming one. Graphing the projects' budget overruns reveals a "fat tail" — a large number of gigantic overages. Fully one in six of the projects we studied was a black swan, with a cost overrun of 200%, on average, and a schedule overrun of almost 70%. This highlights the true pitfall of IT change initiatives: It's not that they're particularly prone to high cost overruns *on average*, as management consultants and academic studies have previously suggested. It's that an unusually large proportion of them incur massive overages — that is, there are a disproportionate number of Black Swans. By focusing on averages instead of the more damaging outliers, most managers and consultants have been missing the real problem.

Some of the pitfalls of tech projects are old ones. More than a decade ago, for example, Hershey's shift to a new order-taking and fulfillment system prevented the company from shipping $100 million worth of candy in time for Halloween, causing an 18.6% drop in quarterly earnings. Our research suggests that such problems are now occurring systematically. The biggest ones typically arise in companies facing serious difficulties — eroding margins, rising cost pressures, demanding debt servicing, and so on — which an out-of-control tech project can fatally compound. Kmart was already losing its competitive position to Walmart and Target when it began a $1.4 billion IT modernization project in 2000. By 2001 it had realized that the new system was so highly customized that maintenance would be prohibitively expensive. So it launched a $600 million project to update its supply chain management software. That effort went off the rails in 2002, and the two projects

contributed to Kmart's decision to file for bankruptcy that year. The company later merged with Sears Holdings, shedding more than 600 stores and 67,000 employees.

Other countries, too, have seen companies fail as the result of flawed technology projects. In 2006, for instance, Auto Windscreens was the second-largest automobile glass company in the UK, with 1,100 employees and £63 million in revenue. Unsatisfied with its financial IT system, the company migrated its order management from Oracle to Metrix and started to implement a Microsoft ERP system. In the fourth quarter of 2010, a combination of falling sales, inventory management problems, and spending on the IT project forced it into bankruptcy. Just a few years earlier the German company Toll Collect — a consortium of DaimlerChrysler, Deutsche Telekom, and Cofiroute of France — suffered its own debacle while implementing technology designed to help collect tolls from heavy trucks on German roadways. The developers struggled to combine the different software systems, and in the end the project cost the government more than $10 billion in lost revenue, according to one estimate. "Toll Collect" became a popular byword among Germans for the woes of their economy.

Software is now an integral part of numerous products — think of the complex software systems in cars and consumer appliances — but the engineers and managers who are in charge of product development too often have a limited understanding of how to implement the technology component. That was the case at Airbus, whose A380 was conceived to take full advantage of cutting-edge technology: Its original design, finalized in 2001, called for more than 300 miles of wiring, 98,000 cables, and 40,000 connectors per aircraft. Partway through the project the global product development team learned that the German and Spanish facilities were using an older version of the product development software than the British and French facilities were; configuration problems inevitably ensued. In 2005 Airbus announced a six-month delay in its first delivery. The following year it announced another six-month delay, causing a 26% drop in share price and prompting several high-profile resignations. By 2010 the company still had not caught up with production plans, and the continuing problems with the A380 had led to further financial losses and reputational damage.

**Avoiding Black Swans**

Any company that is contemplating a large technology project should take a stress test designed to assess its readiness. Leaders should ask themselves two key questions as part of IT black swan management: First, is the company strong enough to absorb the hit if its biggest technology project goes over budget by 400% or more and if only 25% to 50% of the projected benefits are realized? Second, can the company take the hit if 15% of its medium-sized tech projects (not the ones that get all the executive attention but the secondary ones that are often overlooked) exceed cost estimates by 200%? These numbers may seem comfortably improbable, but, as our research shows, they apply with uncomfortable frequency.

Even if their companies pass the stress test, smart managers take other steps to avoid IT black swans. They break big projects down into ones of limited size, complexity, and duration; recognize and make contingency plans to deal with unavoidable risks; and avail themselves of the best possible forecasting techniques — for example, "reference class forecasting," a method based on the Nobel Prize – winning work of Daniel Kahneman and Amos Tversky. These techniques, which take into account the outcomes of similar projects conducted in other organizations, are now widely used in business, government, and consulting and have become mandatory for big public projects in the UK and Denmark.

As global companies become even more reliant on analytics and data to drive good decision making, periodic overhauls of their technology systems are inevitable. But the risks involved can be profound, and avoiding them requires top managers' careful attention.

**<INSET> Success Story: How One Company Nailed a Tricky IT Project**

In April 2006 emirates Bank decided to revamp parts of its core banking system. After 12 months of planning, managers kicked off the project. They had two main objectives: to avoid mission creep and to go live as soon as possible. During the summer of 2007, however, the bank announced a merger with the national Bank of Dubai, forming Emirates NBD. This immediately made the already-complex project much more daunting: the system now needed to work for both banks — and it had to be ready in 18 months. In addition, it was to be rolled out in a "big bang": all the components — branch computers, ATMs, online banking, and call centers — would be switched to the new system simultaneously. the potential for going way over budget was all too real.  But by the time the project was completed, in November 2009, the schedule had slipped by only 7%, and costs had exceeded the initial estimate by only 18% — even though the merger had doubled the project's size. In a field where massive overruns are common, that's a spectacular achievement.   The project leaders took several key steps. They:

1. Stuck to the schedule, even after the merger
2. Resisted changes to the project's scope
3. Broke the project into discrete modules
4. Assembled the right team, including It experts from both companies, outside experts, and vendors
5. Prevented turnover among team members
6. Framed the initiative as a business endeavor, not a technical one
7. Focused on a single target, "readiness to go live," measuring every activity against it.